\documentclass[prb,twocolumn,showpacs,superscriptaddress,amsmath]{revtex4}
\usepackage{graphicx}
\usepackage{amssymb}
\usepackage{citesort}

\begin{document}
\title{Temperature-driven transition from the Wigner Crystal to the
Bond-Charge-Density Wave in the Quasi-One-Dimensional Quarter-Filled band}
\author{R.T. Clay}
\affiliation{Department of Physics and Astronomy and HPC$^2$ Center for 
Computational Sciences, Mississippi State University, Mississippi State MS 39762}
\author{R.P. Hardikar}
\affiliation{Department of Physics and Astronomy and HPC$^2$ Center for 
Computational Sciences, Mississippi State University, Mississippi State MS 39762}
\author{S. Mazumdar}
\affiliation{ Department of Physics, University of Arizona
Tucson, AZ 85721}
\date{\today}
\begin{abstract}
It is known that within the interacting electron model Hamiltonian for
the one-dimensional $\frac{1}{4}$-filled band, the singlet ground
state is a Wigner crystal only if the nearest neighbor
electron-electron repulsion is larger than a critical value. We show
that this critical nearest neighbor Coulomb interaction is different
for each spin subspace, with the critical value decreasing with
increasing spin. As a consequence, with the lowering of temperature,
there can occur a transition from a Wigner crystal charge-ordered
state to a spin-Peierls state that is a Bond-Charge-Density Wave with
charge occupancies different from the Wigner crystal.  This transition
is possible because spin excitations from the spin-Peierls state in
the $\frac{1}{4}$-filled band are necessarily accompanied by changes
in site charge densities. We apply our theory to the
$\frac{1}{4}$-filled band quasi-one-dimensional organic
charge-transfer solids in general and to 2:1
tetramethyltetrathiafulvalene (TMTTF) and
tetramethyltetraselenafulvalene (TMTSF) cationic salts in
particular. We believe that many recent experiments strongly indicate
the Wigner crystal to Bond-Charge-Density Wave transition in several
members of the TMTTF family. We explain the occurrence of two
different antiferromagnetic phases but a single spin-Peierls state in
the generic phase diagram for the 2:1 cationic solids. The
antiferromagnetic phases can have either the Wigner crystal or the
Bond-Charge-Spin-Density Wave charge occupancies. The spin-Peierls
state is always a Bond-Charge-Density Wave.
\end{abstract}

\maketitle

\section{Introduction}

Spatial broken symmetries in the quasi-one-dimensional (quasi-1D)
$\frac{1}{4}$-filled organic charge-transfer solids (CTS) have been of
strong experimental
\cite{Pouget96a,Dumoulin96a,Kanoda,Nad06a,Monceau01a,Nad05a,Chow00a,Zamborszky02a,Yu04a,Foury-Leylekian04a,Takahashi06a,Nakamura03a,Fujiyama04a,Nakamura06a,Fujiyama06a,Nakamura07a,Dumm}
and theoretical
\cite{Ung94a,Penc94a,Seo97a,Mazumdar99a,Riera00a,Riera01a,Shibata01a,Clay03a,Seo06a}
interest. The broken symmetry states include charge order (hereafter
CO, this is usually accompanied by intramolecular distortions),
intermolecular lattice distortions (hereafter bond order wave or BOW),
antiferromagnetism (AFM) and spin-Peierls (SP) order.  Multiple
orderings may compete or even coexist simultaneously.  Interestingly,
these unconventional insulating states in the CTS are often proximate
to superconductivity \cite{Ishiguro}, the mechanism of which has
remained perplexing after intensive investigations over several
decades. Unconventional behavior at or near $\frac{1}{4}$-filling has
also been observed in the quasi-two-dimensional organic CTS with
higher superconducting critical temperatures
\cite{Mori98b,Takahashi06a,Kawamoto04a,Miyagawa, Powell06a}, sodium
cobaltate \cite{Pedrini05a,Lee06a,Choy07a} and oxides of titanium
\cite{Lakkis76a} and vanadium \cite{Isobe96a,Yamauchi02a}.
 \begin{figure}[tb]
\centerline{\resizebox{2.7in}{!}{\includegraphics{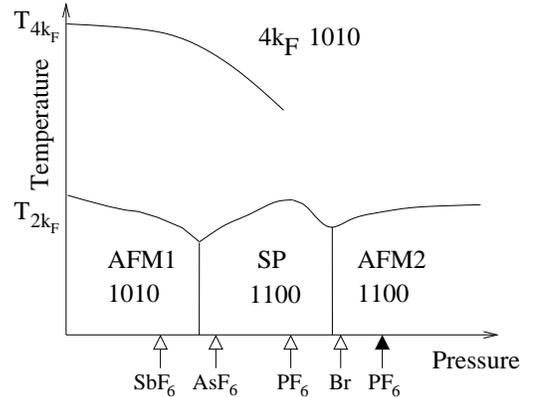}}}
\caption{Schematic of the proposed T-vs.-P phase diagram for
(TMTCF)$_2$X, where C=T or S, along with the charge occupancies of the
sites in the low T phases as determined in this work. The P-axis
reflects the extent of interchain coupling.  Open (filled) arrows
indicate the ambient pressure locations of TMTTF (TMTSF) salts.}
\label{fig-phasediag}
\end{figure}

In spite of extensive research on 1D instabilities in the CTS,
detailed comparisons of theory and experiments remain difficult.
Strong electron-electron (e-e) Coulomb interactions in these systems
make calculations particularly challenging, and with few exceptions
\cite{Yoshioka06a,Seo07a} existing theoretical discussions of broken
symmetries in the interacting $\frac{1}{4}$-filled band have been
limited to the {\it ground} state
\cite{Ung94a,Mazumdar99a,Clay03a,Penc94a,Seo97a,Seo06a,Riera00a,Riera01a,Shibata01a}.
This leaves a number of important questions unresolved, as we point
out below.

Quasi-1D CTS undergo two distinct phase transitions as the temperature
(hereafter T) is reduced \cite{footnote}.  The 4k$_F$ transition at
higher T involves charge degrees of freedom (T$_{4k_F} \sim$ 100 K),
with the semiconducting state below T$_{4k_F}$ exhibiting either a
dimerized CO or a dimerized BOW. Charges alternate as 0.5 + $\epsilon$
and 0.5 -- $\epsilon$ on the molecules along the stack in the
dimerized CO.  The site charge occupancy in the dimerized CO state is
commonly written as $\cdots$1010$\cdots$ (with `1' and `0' denoting
charge-rich and charge-poor sites), and this state is also referred to
as the Wigner crystal. The 4k$_F$ BOW has alternating intermolecular
bond strengths, but site charges are uniformly 0.5. It is generally
accepted that the dimer CO (dimer BOW) is obtained for strong (weak)
intermolecular Coulomb interactions (the intramolecular Coulomb
interaction can be large in either case, see Section III).  At T $<
T_{2k_F} \sim$ 10 -- 20 K, there occurs a second transition in the
CTS, involving spin degrees of freedom, to either an SP or an AFM
state. Importantly, the SP and the AFM states both can be derived from
the dimer CO or the dimer BOW. Coexistence of the SP state with the
Wigner crystal would require that the SP state has the structure
$1=0=1\cdots0\cdots1$, with singlet ``bonds'' alternating in strength
between the charge-rich sites of the Wigner crystal.  Similarly,
coexisting AFM and Wigner crystal would imply site charge-spin
occupancies $\uparrow 0 \downarrow 0$. The occurrence of both have
been suggested in the literature
\cite{Seo97a,Seo06a,Seo07a,Yoshioka06a}.  The SP state can also be
obtained from dimerization of the dimer 4k$_F$ BOW, in which case
there occurs a spontaneous transition from the uniform charge-density
state to a coexisting bond-charge-density wave (BCDW)
\cite{Ung94a,Clay03a,Riera00a} $1-1=0\cdots0$.  The unit cells here
are dimers with site occupancies `1' and `0' or `0' and `1', with
strong and weak interunit 1--1 and 0$\cdots$0 bonds, respectively.
The AFM state with the same site occupancies $\cdots$1100$\cdots$ is
referred to as the bond-charge-spin density wave (BCSDW)
\cite{Mazumdar99a} and is denoted as $\uparrow \downarrow 0 0$.

The above characterizations of the different states and the related
bond, charge and spin patterns are largely based on ground state
calculations. The mechanism of the transition from the 4k$_F$ state to
the 2k$_F$ state lies outside the scope of such theories.  Questions
that remain unresolved as a consequence include: (i) Does the nature
of the 4k$_F$ state (charge versus bond dimerized) predetermine the
site charge occupancies of the 2k$_F$ state?  This is assumed in many
of the published works.  (ii) What determines the nature of the 2k$_F$
state (AFM versus SP)?  (iii) How does one understand the occurrence
of two different AFM phases (hereafter AFM1 and AFM2) straddling a
single SP phase in the proposed T vs. P (where P is pressure) phase
diagram (see Fig.~\ref{fig-phasediag}) \cite{Yu04a} for the cationic
$\frac{1}{4}$-filled band quasi-1D CTS? (iv) What is the nature of the
intermediate T states between the 4k$_F$ dimerized state and the
2k$_F$ tetramerized state? As we point out in the next section, where
we present a brief review of the experimental results, answers to
these questions are crucial for a deeper understanding of the
underlying physics of the $\frac{1}{4}$-filled band CTS.

In the present paper we report the results of our calculations of
T-dependent behavior within theoretical models incorporating both e-e
and electron-phonon (e-p) interactions to answer precisely the above
questions. One key result of our work is as follows: in between the
strong and weak intermolecular Coulomb interaction parameter regimes,
there exists a third parameter regime within which the intermolecular
Coulomb interactions are intermediate, and within which there can
occur a novel transition from a Wigner crystal CO state to a BCDW SP
state as the T is lowered.  Thus the charge or bond ordering in the
4k$_F$ phase {\it does not} necessarily decide the same in the 2k$_F$
state.  For realistic intramolecular Coulomb interactions (Hubbard
$U$), we show that the width of this intermediate parameter regime is
comparable to the strong and weak intersite interaction regimes. We
believe that our results are directly applicable to the family of the
cationic CTS (TMTTF)$_2$X, where a redistribution of charge upon
entering the SP state from the CO state has been observed
\cite{Fujiyama06a,Nakamura07a} in X=AsF$_6$ and PF$_6$.  A natural
explanation of this redistribution emerges within our theory.  The SP
state within our theory is unique and has the BCDW charge occupancy,
while the two AFM regions in the phase diagram of
Fig.~\ref{fig-phasediag} have different site occupancies.  Our theory
therefore provides a simple diagnostic to determine the pattern of CO
coexisting with low-temperature magnetic states in the
$\frac{1}{4}$-filled CTS.

In addition to the above theoretical results directly pertaining to
the quasi-1D CTS, our work gives new insight to excitations from a SP
ground state in a non-half-filled band. In the case of the usual SP
transition within the $\frac{1}{2}$-filled band, the SP state is
bond-dimerized at T = 0 and has uniform bonds at T $> T_{2k_F}$. The
site charges are uniform at all T. This is in contrast to the
$\frac{1}{4}$-filled band, where the SP state at T = 0 is bond and
charge-tetramerized and the T $> T_{2k_F}$ state is dimerized as
opposed to being uniform. Furthermore, we show that the high T phase
here can be either charge- or bond-dimerized, starting from the same
low T state. This clearly requires {\it two different kinds of spin
excitations} in the $\frac{1}{4}$-filled band.  We demonstrate that
spin excitations from the SP state in the $\frac{1}{4}$-filled band
can lead to two different kinds of defects in the background BCDW.

In the next section we present a brief yet detailed summary of
relevant experimental results in the quasi-1D CTS. The scope of this
summary makes the need for having T-dependent theory clear. Following
this, in Section III we present our theoretical model along with
conjectures based on physical intuitive pictures. In Section IV we
substantiate these conjectures with accurate quantum Monte Carlo (QMC)
and exact diagonalization (ED) numerical calculations. Finally in
Section V we compare our theoretical results and experiments, and
present our conclusions.

\section{Review of Experimental Results}

Examples of both CO and BOW broken symmetry at T $<$ T$_{4k_F}$ are
found in the $\frac{1}{4}$-filled CTS. The 4k$_F$ phase in the anionic
1:2 CTS is commonly bond-dimerized.  The most well known example is
MEM(TCNQ)$_2$, which undergoes a metal-insulator transition
accompanied with bond-dimerization at 335 K \cite{Visser83a}. Site
charges are uniform in this 4k$_F$ phase.  The 2k$_F$ phase in the
TCNQ-based systems is universally SP and not AFM.  The SP transition
in MEM(TCNQ)$_2$ occurs below T$_{2k_F}$ = 19 K, and low T neutron
diffraction measurements of deuterated samples \cite{Visser83a} have
established that the bond tetramerization is accompanied by 2k$_F$ CO
$\cdots$1100$\cdots$ .  X-ray \cite{Kobayashi70a,Filhol84a} and
neutron diffraction \cite{Filhol80a} experiments have confirmed a
similar low T phase in TEA(TCNQ)$_2$. We will not discuss these
further in the present paper, as they are well described within our
previous work \cite{Ung94a,Clay03a}.  We will, however, argue that the
SP ground state in (DMe-DCNQI)$_2$Ag \cite{Nakazawa03a} (as opposed to
AFM) indicates $\cdots$1100$\cdots$ CO in this.

The cationic (TMTCF)$_2$X, C= S and Se, exhibit more variety,
presumably because the counterions affect packing as well as site
energies in the cation stack. Differences between systems with
centrosymmetric and noncentrosymmetric anions are also observed.
Their overall behavior is summarized in Fig.~\ref{fig-phasediag},
where as is customary pressure P can also imply larger interchain
coupling. We have indicated schematically the possible locations of
different materials on the phase diagram. The most significant aspect
of the phase diagram is the occurrence of two distinct
antiferromagnetic phases, AFM1 and AFM2, straddling a single SP phase.

Most TMTTF lie near the low P region of the phase diagram and are
insulating already at or near room temperature because of charge
localization, which is due to the intrinsic dimerization along the
cationic stacks \cite{Pouget96a,Nad06a,Takahashi06a}.  CO at
intermediate temperatures T$_{CO}$ has been found in dielectric
permittivity \cite{Nad06a}, NMR \cite{Chow00a}, and ESR
\cite{Coulon07a} experiments on materials near the low and
intermediate P end.  Although the pattern of the CO has not been
determined directly, the observation of ferroelectric behavior below
T$_{CO}$ is consistent with $\cdots$1010$\cdots$ type CO in this
region \cite{Monceau01a,Riera01a}.  With further lowering of T, most
(TMTTF)$_2$X undergo transitions to the AFM1 or SP phase (with X = Br
a possible exception, see below). X = SbF$_6$ at low T lies in the
AFM1 region \cite{Yu04a}, with a very high T$_{CO}$ and relatively low
Neel temperature T$_N$ = 8 K. As the schematic phase diagram
indicates, pressure suppresses both T$_{CO}$ and T$_N$ in this
region. For P $>$ 0.5 GPa, (TMTTF)$_2$SbF$_6$ undergoes a transition
from the AFM1 to the SP phase \cite{Yu04a}, the details of which are
not completely understood; any charge disproportionation in the SP
phase is small \cite{Yu04a}.  (TMTTF)$_2$ReO$_4$ also has a relatively
high T$_{CO}$ = 225 K, but the low T phase here, reached following an
anion-ordering transition is spin singlet \cite{Nakamura06a}. Nakamura
{\it et al.} have suggested, based on NMR experiments, that the CO
involves the Wigner crystal state, but the low T state is the
$\cdots$1100$\cdots$ BCDW \cite{Nakamura06a}. Further along the P axis
lie X = AsF$_6$ and PF$_6$, where T$_{CO}$ are reduced to 100 K and 65
K, respectively \cite{Chow00a}. The low T phase in both cases is now
SP.  Neutron scattering experiments on (TMTTF)$_2$PF$_6$ have found
that the lattice distortion in the SP state is the expected 2k$_F$ BOW
distortion, but that the amplitude of the lattice distortion is much
smaller \cite{Foury-Leylekian04a} than that found in other organic SP
materials such as MEM(TCNQ)$_2$. The exact pattern of the BOW has not
been determined yet. Experimental evidence exists that some form of CO
persists in the magnetic phases. For example, the splitting in
vibronic modes below T$_{CO}$ in (TMTTF)$_2$PF$_6$ and
(TMTTF)$_2$AsF$_6$, a signature of charge disproportionation, persists
into the SP phase \cite{Dumm}, indicating coexistence of CO and SP.
At the same time, the high T CO is in competition with the SP ground
state \cite{Zamborszky02a}, as is inferred from the different effects
of pressure on T$_{CO}$ and T$_{SP}$: while pressure reduces T$_{CO}$,
it {\it increases} T$_{SP}$. This is in clear contrast to the effect
of pressure on T$_N$ in X = SbF$_6$. Similarly, deuteration of the
hydrogen atoms of TMTTF increases T$_{CO}$ but decreases
\cite{Nad05a,Furukawa05a} T$_{SP}$. That higher T$_{CO}$ is
accompanied by lower T$_{SP}$ for centrosymmetric X (T$_{SP}$= 16.4 K
in X = PF$_6$ and 11.1 K in X = AsF$_6$) has also been noted
\cite{Pouget06a}. This trend is in obvious agreement with the
occurrence of AFM instead of SP state under ambient pressure in X =
SbF$_6$. Most interestingly, Nakamura {\it et al.} have very recently
observed redistribution of the charges on the TMTTF molecules in
(TMTTF)$_2$AsF$_6$ and (TMTTF)$_2$PF$_6$ as these systems enter the SP
phase from CO states \cite{Fujiyama06a,Nakamura07a}. Charge
disproportionation, if any, in the SP phase is much smaller than in
the CO phase \cite{Fujiyama06a,Nakamura07a}, which is in apparent
agreement with the above observations \cite{Nakamura06a,Yu04a} in X =
ReO$_4$ and SbF$_6$.

The bulk of the (TMTTF)$_2$X therefore lie in the AFM1 and SP regions
of Fig.~\ref{fig-phasediag}.  (TMTSF)$_2$X, in contrast, occupy the
AFM2 region. Coexisting 2k$_F$ CDW and spin-density wave, SDW, with
the {\it same} 2k$_F$ periodicity \cite{Pouget97a,Kagoshima99a} here
is explained naturally as the $\cdots$1100$\cdots$ BCSDW
\cite{Mazumdar99a,Riera00a,Kobayashi97a}.  In contrast to the TMTTF
salts discussed above, charge and magnetic ordering in (TMTTF)$_2$Br
occur almost simultaneously \cite{Fujiyama02a,Coulon07a}.  X-ray
studies of lattice distortions point to similarities with
(TMTSF)$_2$PF$_6$ \cite{Pouget97a}, indicating that (TMTTF)$_2$Br is
also a $\cdots$1100$\cdots$ BCSDW \cite{Mazumdar99a}. We do not
discuss AFM2 region in the present paper, as this can be found in our
earlier work \cite{Mazumdar99a,Clay03a}.

\section{Theoretical Model and Conjectures}

The 1D Hamiltonian we investigate is written as
\begin{subequations}
\begin{eqnarray}
H &=& H_{SSH} + H_{Hol} + H_{ee} \label{eqn-h} \\
H_{SSH} &=& t\sum_i[1+\alpha(a_i^{\dagger}+a_i)](c_{i,\sigma}^{\dagger}c_{i+1,\sigma} + h.c.)\nonumber \\
&+& \hbar \omega_S \sum_i a_i^{\dagger}a_i \\
H_{Hol} &=& g \sum_i (b_i^{\dagger} + b_i)n_i + \hbar \omega_{H} \sum_i b_i^{\dagger}b_i \\
H_{ee} &=& U\sum_i n_{i,\uparrow}n_{i,\downarrow} + V\sum_i n_in_{i+1} \label{eqn-uv}
\end{eqnarray}
\end{subequations}
In the above, $c_{i,\sigma}^{\dagger}$ creates an electron with spin
$\sigma$ ($\uparrow$,$\downarrow$) on molecular site $i$,
$n_{i,\sigma}=c_{i,\sigma}^{\dagger}c_{i,\sigma}$ is the number of
electrons with spin $\sigma$ on site $i$, and
$n_i=\sum_{\sigma}n_{i,\sigma}$. $U$ and $V$ are the on-site and
intersite Coulomb repulsions, and $a_i^{\dagger}$ and $b_i^{\dagger}$
create (dispersionless) Su-Schrieffer-Heeger (SSH) \cite{Su79a} and
Holstein (Hol) \cite{Holstein59a} phonons on the $i$th bond and site
respectively, with frequencies $\omega_S$ and $\omega_H$.  Because the
Peierls instability involves only phonon modes near $q=\pi$, keeping
single dispersionless phonon modes is sufficient for the Peierls
transitions to occur \cite{Sengupta03a,Louis05a}.  Although purely 1D
calculations cannot yield a finite temperature phase transition, as in
all low dimensional theories \cite{Cross79a,Hirsch84a} we anticipate
that the 3D ordering in the real system is principally determined by
the dominant 1D instability.

The above Hamiltonian includes the most important terms necessary to
describe the family of quasi-1D CTS, but ignores nonessential terms
that may be necessary for understanding the detailed behavior of
individual systems. Such nonessential terms include (i) the intrinsic
dimerization that characterizes many (TMTTF)$_2$X, (ii) interaction
between counterions and the carriers on the quasi-1D cations stacks,
(iii) interchain Coulomb interaction, and (iv) interchain
hopping. Inclusion of the intrinsic dimerization will make the Wigner
crystal ground state even less likely \cite{Shibata01a}, and this is
the reason for excluding it. We have verified the conclusions of
reference \onlinecite{Shibata01a} from exact diagonalization
calculations. The inclusion of interactions with counterions may
enhance the Wigner crystal ordering\cite{Monceau01a,Riera01a} for some
(TMTTF)$_2$X. We will discuss this point further below, and argue that
it is important in the AFM1 region of the phase diagram.  The effects
of intrinsic dimerization and counterion interactions can be
reproduced by modifying the $V/|t|$ in our Hamiltonian, and thus these
are not included explicitly. Rather the $V$ in Eq.~(\ref{eqn-h}) should
be considered as the effective $V$ for the quasi-1D CTS. Interchain
hopping, at least in the TMTTF (though not in the TMTSF), are indeed
negligible. The interchain Coulomb interaction can promote the Wigner
crystal within a rectangular lattice framework, but for the realistic
nearly triangular lattice will cause frustration, thereby making the
BCDW state of interest here even more likely. We therefore believe
that Hamiltonian (\ref{eqn-h}) captures the essential physics of the quasi-1D
CTS.

For applications to the quasi-1D CTS we will be interested in the
parameter regime \cite{Ung94a,Shibata01a,Clay03a,Seo06a}
$|t|=0.1-0.25$ eV, $U/|t|=6-8$. The exact value of $V$ is less known,
but since the same cationic molecules of interest as well as other
related molecules (for e.g., HMTTF, HMTSF, etc.) also form quasi-1D
$\frac{1}{2}$-filled band Mott-Hubbard semiconductors with short-range
antiferromagnetic spin correlations \cite{Hawley78a,Torrance80a},
it must be true that $V<\frac{1}{2}U$ (since for $V>\frac{1}{2}U$ the
1D $\frac{1}{2}$-filled band is a CDW \cite{Dixit84a,Hirsch84b}). Two
other known theoretical results now fix the range of $V$ that will be
of interest. First, the ground state of the $\frac{1}{4}$-filled band
within Hamiltonian (\ref{eqn-h}) in the limit of zero e-p coupling is
the Wigner crystal $\cdots$1010$\cdots$ only for sufficiently large
$V>V_c(U)$. Second, $V_c(U \to \infty)=2|t|$, and is larger for finite
$U$ \cite{Penc94a,Shibata01a,Clay03a}. With the known $U/|t|$ and the
above restrictions in place, it is now easily concluded that (i) the
Wigner crystal is obtained in the CTS for a relatively narrow range of
realistic parameters, and (ii) {\it even in such cases the material's
$V$ is barely larger than $V_c(U)$} \cite{Clay03a,Shibata01a}.

We now go beyond the above ground state theories of spatial broken
symmetries to make the following crucial observation: {\it each
different spin subspace of Hamiltonian (\ref{eqn-h}) must have its own
$V_c$ at which the Wigner crystal is formed}.  This conclusion of a
{\it spin-dependent} $V_c=V_c(U,S)$ follows from the comparison of the
ferromagnetic subspace with total spin $S=S_{max}$ and the $S=0$
subspace. The ferromagnetic subspace is equivalent to the
$\frac{1}{2}$-filled spinless fermion band, and therefore
$V_c(U,S_{max})$ is independent of $U$ and exactly 2$|t|$.  The
increase of $V_c(U,S=0)$ with decreasing $U$ \cite{Penc94a,Clay03a} is
then clearly related to the occurrence of doubly occupied and vacant
sites at finite $U$ in the $S=0$ subspace. Since the probability of
double occupancy (for fixed $U$ and $V$) decreases monotonically with
increasing $S$, we can interpolate between the two extreme cases of
$S=0$ and $S=S_{max}$ to conclude $V_c(U,S)>V_c(U,S+1)$. We prove this
explicitly from numerical calculations in the next section.

Our conjecture regarding spin-dependent $V_c(U)$ in turn implies that
there exist three distinct parameter regimes for realistic $U$ and
$V$: (i) $V \leq V(S_{max})=2|t|$, in which case the ground state is
the BCDW with 2k$_F$ CO and the high temperature 4k$_F$ state is a BOW
\cite{Ung94a,Clay03a}; (ii) $V>V_c(U,S=0)$, in which case both the
4k$_F$ and the 2k$_F$ phases have Wigner crystal CO; (iii) and the
intermediate regime $2|t| \leq V \leq V_c(U,S=0)$. Parameter regime
(i) has been discussed in our previous work
\cite{Ung94a,Mazumdar99a,Mazumdar00a,Clay03a}.  This is the case where
the description of the $\frac{1}{4}$-filled band as an effective
$\frac{1}{2}$-filled band is appropriate: the unit cell in the 4k$_F$
state is a dimer of two sites, and the 2k$_F$ transition can be
described as the usual SP dimerization of the dimer lattice. We will
not discuss this parameter region further. Given the $U/|t|$ values
for the CTS, parameter regime (ii) can have rather narrow width.  For
$U/|t| = 6$, $V_c(U,S=0)=3.3|t|$, and there is no value of realistic $V$
for which the ground state is a Wigner crystal. For $U/|t| = 8$,
$V_c(U,S=0) = 3.0|t|$, and the widths of parameter regime (iii), $2|t| \leq
V \leq 3|t|$ and of parameter regime (ii), $3|t| \leq V \leq 4|t|$, are
comparable.

We will discuss parameter regime (ii) only briefly, as the physics of
this region has been discussed by other authors
\cite{Seo97a,Seo06a}. We investigated thoroughly the intermediate
parameter regime (iii), which has not been studied previously.  Within
the intermediate parameter regime we expect a novel transition from
the BCDW in the 2k$_F$ state at low T to a $\cdots$1010$\cdots$ CO in
the 4k$_F$ phase at high T that has not been discussed in the
theoretical literature.

Our observation regarding the T-dependent behavior of these CTS
follows from the more general observation that thermodynamic behavior
depends on the free energy and the partition function. For the strong
e-e interactions of interest here, thermodynamics of 1D systems at
temperatures of interest is determined almost entirely by spin
excitations.  Since multiplicities of spin states increase with the
total spin S, the partition function is dominated by high (low) spin
states with large (small) multiplicities at high (low) temperatures.
While at T=0 such a system must be a BCDW for $V<V_c(U,S=0)$, as the
temperature is raised, higher and higher spin states begin to dominate
the free energy, until $V$ for the material in question exceeds
$V_c(U,S)$, at which point the charge occupancy reverts to
$\cdots$1010$\cdots$ We demonstrate this explicitly in the next
section.  A charge redistribution is {\it expected} in such a system
at T$_{2k_F}$, as the deviation from the average charge of 0.5 is much
smaller in the BCDW than in the Wigner crystal \cite{Clay03a}.

The above conjecture leads to yet another novel implication.  The
ground state for both weak and intermediate intersite Coulomb
interaction parameters are the BCDW, even as the 4k$_F$ phases are
different in the two cases: BOW in the former and CO in the
latter. This necessarily requires the existence of {\it two different
kinds of spin excitations from the BCDW}. Recall that within the
standard theory of the SP transition in the $\frac{1}{2}$-filled band
\cite{Cross79a,Sorensen98a,Yu00a}, thermal excitations from the T = 0
ground state generates spin excitations with bond-alternation domain
walls (solitons), with the phase of bond alternation in between the
solitons being opposite to that outside the solitons. Progressive
increase in T generates more and more solitons with reversed bond
alternation phases, until overlaps between the two phases of bond
alternations lead ultimately to the uniform state. A key difference
between the $\frac{1}{2}$-filled and $\frac{1}{4}$-filled SP states is
that site charge occupancies in spin excitations in the former
continue to be uniform, while they are necessarily nonuniform in the
latter case, as we show in the next section. We demonstrate that
defect centers with two distinct charge occupancies (that we will term
as type I and II), depending upon the actual value of $V$, are
possible in the $\frac{1}{4}$-filled band. Preponderance of one or
another type of defects generates the distinct 4k$_F$ states.

\section{Results}

We present in this section the results of QMC investigations of
Eq.~(\ref{eqn-h}), for both zero and nonzero e-p couplings, and of ED
studies of the adiabatic (semiclassical) limit of
Eq.~(\ref{eqn-h}). Using QMC techniques, we demonstrate explicitly the
spin-dependence of V$_c$(U), as well as the transition from the Wigner
crystal to the BCDW for the intermediate parameter regime (iii). The
ED studies demonstrate the exotic nature of spin excitations from the
BCDW ground state. In what follows, all quantities are expressed in
units of $|t|$ ($|t|$=1).

The QMC method we use is the Stochastic Series Expansion (SSE) method
using the directed loop update for the electron degrees of freedom
\cite{Syljuasen02a}.  For 1D fermions with nearest-neighbor hopping
SSE provides statistically exact results with no systematic errors.
While the SSE directed-loop method is grand canonical (with
fluctuating particle density), we restrict measurements to only the
$\frac{1}{4}$-filled density sector to obtain results in the canonical
ensemble.  Quantum phonons are treated within SSE by directly adding
the bosonic phonon creation and annihilation operators in
Eq.~(\ref{eqn-h}) to the series expansion \cite{Hardikar07a}.  An upper
limit in the phonon spectrum must be imposed, but can be set
arbitrarily large to avoid any systematic errors \cite{Hardikar07a}.
For the results shown below we used a cutoff of 100 SSH phonons per
bond and either 30 (for $g=0.5$) or 50 (for $g=0.75$) Holstein phonons
per site.

The observables we calculate within SSE are the standard wave-vector
dependent charge structure factor $S_\rho(q)$, defined as,
\begin{equation}
S_\rho(q)=\frac{1}{N} \sum_{j,k} e^{iq(j-k)} \langle O^\rho_j O^\rho_k \rangle
\label{eqn:sfac}
\end{equation}
 and charge and bond-order susceptibilities $\chi_\rho(q)$ and $\chi_B(q)$, defined as,
\begin{equation}
  \chi_x(q) = \frac{1}{N} \sum_{j,k} e^{iq(j-k)} \int_{0}^{\beta} d\tau
 \langle O^x_j(\tau)O^x_k(0)\rangle
\label{eqn:susceptibility}
\end{equation}
In Eqs.~(\ref{eqn:sfac}) and (\ref{eqn:susceptibility}) $N$ is the
number of lattice sites, $O^\rho_j = n_{j,\uparrow} +
n_{j,\downarrow}$, $O^B_j = \sum_\sigma
(c^\dagger_{j+1,\sigma}c_{j,\sigma}+h.c.)$, and $\beta$ is the inverse
temperature in units of $t$.

The presence of CO or BOW can be detected by the divergence of the
2k$_F$ or 4k$_F$ charge or bond-order susceptibility as a function of increasing
system size. Strictly speaking, in a purely 1D model these functions 
diverge only at T = 0; as already explained above, we make the reasonable
assumption \cite{Hirsch84a} that in the presence of realistic inter-chain couplings transitions
involving charge or bond-order instabilities, as determined
by the dominant susceptibility, occur at finite T.

\subsection{Spin-dependent V$_c$(U)}

\begin{figure}[tb]
\centerline{\resizebox{3.0in}{!}{\includegraphics{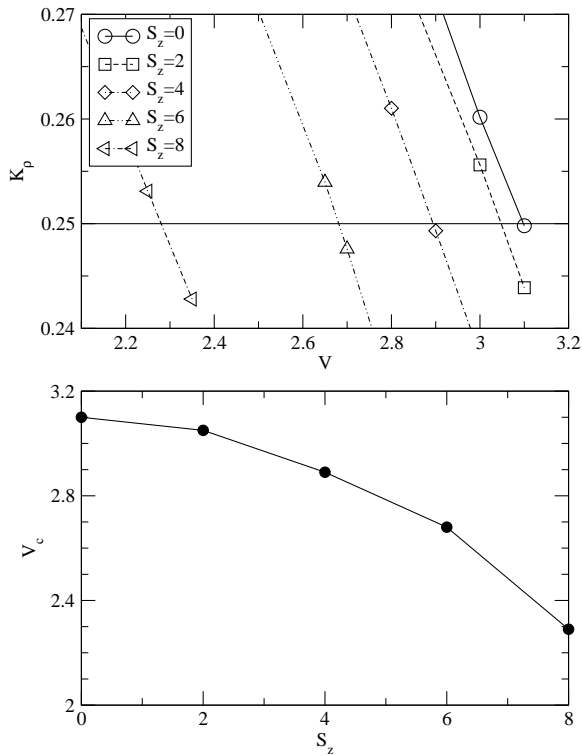}}}
\caption{$V_c$ for Eq.~(\ref{eqn-h}) in the limit of zero e-p
interactions ($\alpha=g=0$) as a function of $S_z$. Results are for a
$N$=32 site periodic ring with $U=8$. For $N=32$ $S_z$=8 corresponds to
the fully polarized (spinless fermion) limit. 
(a) Luttinger Liquid exponent $K_\rho$ as a function of $V$.
$K_\rho=\frac{1}{4}$ determines the boundary for the $\cdots$1010$\cdots$ CO
phase.
 (b) $V_c$ plotted vs. $S_z$.}
\label{fig-vc}
\end{figure}
We first present computational results within Hamiltonian (\ref{eqn-h})
in the absence of e-p coupling to demonstrate that the $V_c$(U) at
which the Wigner crystal order is established in the lowest state of a
given spin subspace decreases with increasing spin $S$.  Our
computational approach conserves the total z-component of spin $S_z$
and not the total $S$. Since the Lieb-Mattis theorem \cite{Lieb62a}
$E(S)<E(S+1)$, where $E(S)$ is the energy of the lowest state in the
spin subspace $S$, applies to the 1D Hamiltonian ~(\ref{eqn-h}), and
since in the absence of a magnetic field all $S_z$ states for a given
$S$ are degenerate, our results for the lowest state within each
different $S_z$ must pertain to $S=S_z$.

To determine $V_c(U,S)$ we use the fact that the purely electronic model is
a Luttinger liquid (LL) for $V<V_c$ with correlation functions
determined by a single exponent $K_\rho$ (see Reference
\onlinecite{Voit95a} for a review). 
The Wigner crystal state is reached when $K_\rho=\frac{1}{4}$.
The exponent $K_\rho$ may be
calculated from the long-wavelength limit of $S_\rho(q)$
\cite{Clay99a}. 
In Fig.~\ref{fig-vc}(a) we have plotted our calculated $K_\rho$
for $U=8$ as a function of $V$ for different $S_z$ sectors. The temperature
chosen is small enough ($\beta=2N$) that in all cases the results 
correspond to the lowest state within a given $S_z$. In
Fig.~\ref{fig-vc}(b) we have plotted our calculated $V_c(U=8)$, as
obtained from Fig.~\ref{fig-vc}(a), as a function of $S_z$. 
$V_c$ is largest for $S_z=0$ and decreases with increasing $S_z$, in
agreement with the conjecture of Section III.  Importantly, the
calculated $V_c$ for $S_z$ = 8 is close to the correct limiting value
of 2, indicating the validity of our approach.
We have not performed any finite-size scaling in Fig.~\ref{fig-vc}, which accounts for the
the slight deviation from the exact value of 2.

\subsection{T-dependent susceptibilities}
\label{sect:qmc}

\begin{figure}[tb]
\centerline{\resizebox{3.0in}{!}{\includegraphics{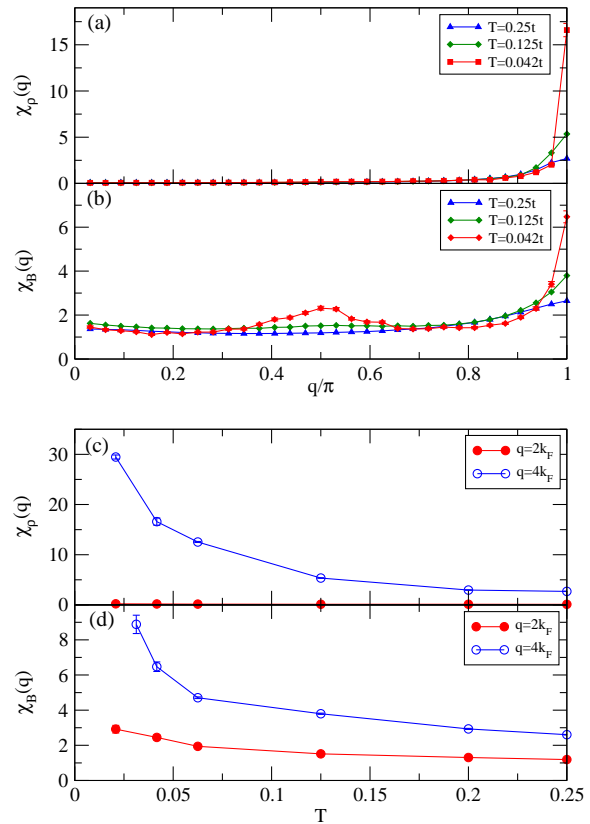}}}
\caption{(color online) QMC results for the temperature-dependent charge
susceptibilities for a N=64 site periodic ring with $U=8$, $V=2.75$,
$\alpha=0.15$, $\omega_S=0.1$, $g=0.5$, and $\omega_H=0.5$.  (a) and (b)
Wavevector-dependent charge and bond-order susceptibilities. (c) 2k$_F$ and 4k$_F$
charge susceptibilities as a function of temperature. (d) 2k$_F$ and
4k$_F$ bond-order susceptibilities as a function of temperature. 
If error bars are not shown, 
statistical error bars are smaller than the symbol
sizes. Lines are guides to the eye.}
\label{fig-v2.75-a.15}
\end{figure}
\begin{figure}[tb]
\centerline{\resizebox{3.0in}{!}{\includegraphics{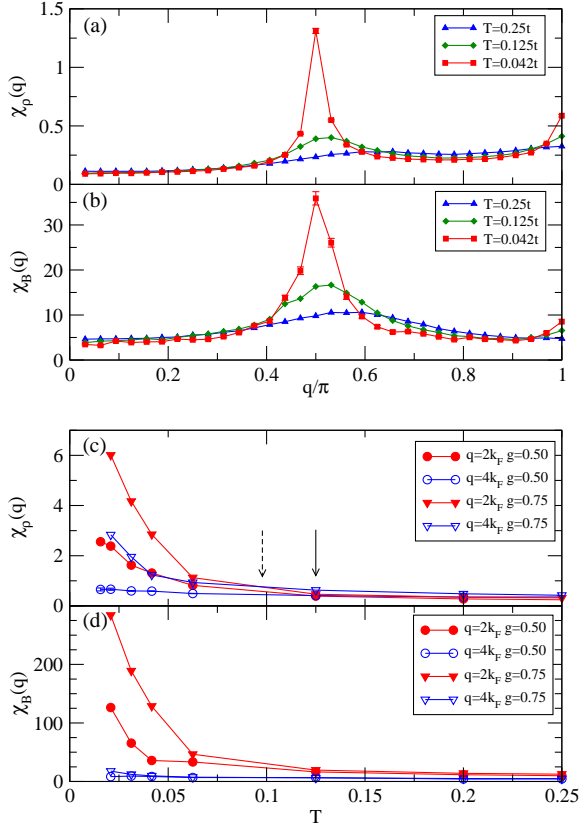}}}
\caption{(color online) Same as Fig.~\ref{fig-v2.75-a.15}, but with
parameters $U=8$, $V=2.25$, $\alpha=0.27$, $\omega_S=0.1$, and
$\omega_H=0.5$. In panels (a) and (b), data are for $g=0.50$ only. In
panels (c) and (d), data for both $g=0.50$ and $g=0.75$ are
shown. Arrows indicate temperature where
$\chi_\rho(2k_F)=\chi_\rho(4k_F)$ (solid and broken arrows correspond
to $g=0.50$ and 0.75, respectively.)}
\label{fig-v2.25-a.27}
\end{figure}
\begin{figure}[tb]
\centerline{\resizebox{3.0in}{!}{\includegraphics{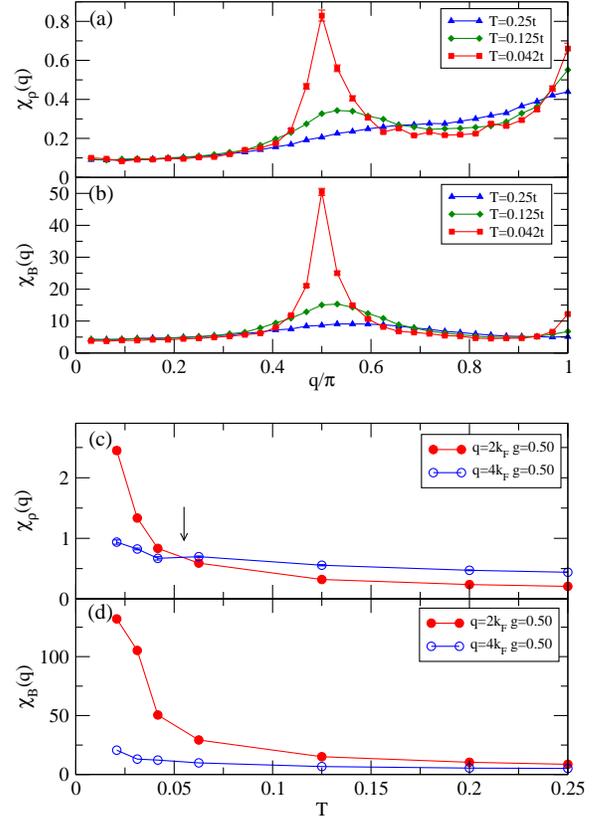}}}
\caption{(color online) Same as Fig.~\ref{fig-v2.75-a.15}, but with
parameters $U=8$, $V=2.75$, $\alpha=0.27$, $\omega_S=0.1$, $g=0.5$,
and $\omega_H=0.5$.  Arrow indicates temperature where
$\chi_\rho(2k_F)=\chi_\rho(4k_F)$.}
\label{fig-v2.75-a.27}
\end{figure}
\begin{figure}[tb]
\centerline{\resizebox{3.0in}{!}{\includegraphics{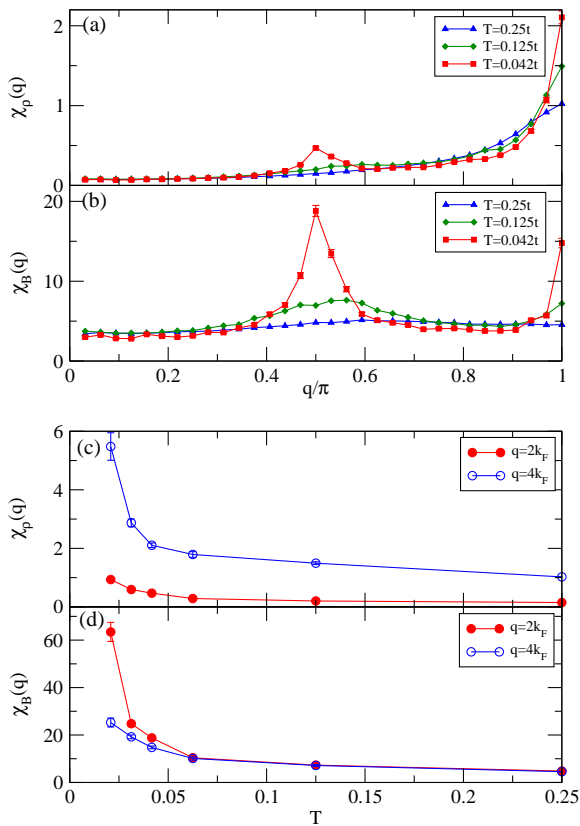}}}
\caption{(color online) Same as Fig.~\ref{fig-v2.75-a.15}, but with 
parameters $U=8$, $V=3.5$,
$\alpha=0.24$, $\omega_S=0.1$, $g=0.5$, and $\omega_H=0.5$.}
\label{fig-v3.5-a.24}
\end{figure}
We next present the results of QMC calculations within the full
Hamiltonian (\ref{eqn-h}).  To reproduce correct relative energy
scales of intra- and intermolecular phonon modes in the CTS, we choose
$\omega_{H}>\omega_{S}$, specifically $\omega_{H}=0.5$ and
$\omega_{S}=0.1$ in our calculations. Small deviations from these
values do not make any significant difference.  In all cases we have
chosen the electron-molecular vibration coupling $g$ larger than the
coupling between electrons and the SSH phonons $\alpha$, thereby
deliberately enhancing the likelihood of the Wigner crystal CO.  We
report results only for intermediate and strong weak intersite Coulomb
interactions; the weak interaction regime $V<2$ has been discussed
extensively in our previous work \cite{Clay03a}.

\subsubsection{$2 \leq V \leq V_c(U,S=0$)}

Our results are summarized in
Figs.~\ref{fig-v2.75-a.15}--\ref{fig-v2.75-a.27}, where we report
results for two different $V$ of intermediate strength and several
different e-p couplings.  Fig.~\ref{fig-v2.75-a.15} first shows
results for relatively weak SSH coupling $\alpha$.  The charge
susceptibility is dominated by a peak at 4k$_F=\pi$
(Fig.~\ref{fig-v2.75-a.15}(a)).  Figs.~\ref{fig-v2.75-a.15}(c) and (d)
show the T-dependence of the 2k$_F$ as well as 4k$_F$ charge and bond
susceptibility.  For $V<V_c(U,S=0)$ the 4k$_F$ charge susceptibility
does not diverge with system size, and the purely 1D system remains a
LL with no long range CO at zero temperature. The dominance of
$\chi_\rho(4k_F)$ over $\chi_\rho(2k_F)$ suggests, however, that
$\cdots$1010$\cdots$ CO will likely occur in the 3D system, especially
if this order is further enhanced due to interactions with the
counterions \cite{Yu04a}.  We have plotted the bond susceptibilities
$\chi_B(2k_F)$ and $\chi_B(4k_F)$ in Fig.~\ref{fig-v2.75-a.15}(d). A
SP transition requires that $\chi_B(2k_F)$ diverges as $T \to
0$. $\chi_B(2k_F)$ weaker than $\chi_B(4k_F)$ at low T indicates that
the SP order is absent in the present case with weak SSH e-p
coupling. This result is in agreement with earlier result
\cite{Sengupta03a} that the SP order is obtained only above a critical
$\alpha_c$ ($\alpha_c$ may be smaller for the infinite system than
found in our calculation).  The most likely scenario with the present
parameters is the persistence of the $\cdots$1010$\cdots$ CO to the
lowest T with no SP transition. These parameters, along with
counterion interactions, could then describe the (TMTTF)$_2$X
materials with an AFM ground state \cite{Yu04a}.

In Figs.~\ref{fig-v2.25-a.27} and Fig.~\ref{fig-v2.75-a.27} we show
our results for larger SSH e-p coupling $\alpha$ and two different
intersite Coulomb interaction $V$=2.25 and 2.75.  The calculations of
Fig.~\ref{fig-v2.25-a.27} were done for two different Holstein e-p
couplings $g$. For both the $V$ parameters, $\chi_\rho(4k_F)$
dominates at high T but $\chi_\rho(2k_F)$ is stronger at low T in both
Fig.~\ref{fig-v2.25-a.27}(c) and Fig.~\ref{fig-v2.75-a.27}(c).  The
crossing between the two susceptibilities is clear indication that in
the intermediate parameter regime, as T is lowered the 2k$_F$ CDW
instability dominates over the 4k$_F$ CO.

The rise in $\chi_\rho(2k_F)$ at low T is accompanied by a steep rise
in $\chi_B(2k_F)$ in both cases (see Fig.~\ref{fig-v2.25-a.27}(d) and
Fig.~\ref{fig-v2.75-a.27}(d)). Importantly, unlike in
Fig.~\ref{fig-v2.75-a.15}(d), $\chi_B(2k_F)$ in these cases clearly
dominates over $\chi_B(4k_F)$ by an order of magnitude at low
temperatures.  There is thus a clear signature of the SP instability
for these parameters.  The {\it simultaneous} rise in $\chi_B(2k_F)$
and $\chi_\rho(2k_F)$ indicates that the SP state is the
$\cdots$1100$\cdots$ BCDW.  Comparison of Figs.~\ref{fig-v2.25-a.27}
and \ref{fig-v2.75-a.27} indicates that the effect of larger $V$ is to
decrease the T where the 2k$_F$ and 4k$_F$ susceptibilities
cross. Since larger $V$ would imply larger T$_{CO}$, this result
implies that larger T$_{CO}$ is accompanied by lower T$_{SP}$. Our
calculations are for relatively modest $\alpha < g$.  Larger $\alpha$
(not shown) further strengthens the divergence of 2k$_F$
susceptibilities.

The motivation for performing the calculations of Fig.~4 with multiple
Holstein couplings was to determine whether it is possible to have a
Wigner crystal at low T even for $V<V(U,S=0)$, by simply increasing
$g$. The argument for this would be that in strong coupling,
increasing $g$ amounts to an effective increase in $V$
\cite{Hirsch83a}.  The Holstein coupling cannot be increased
arbitrarily, however, as beyond a certain point, $g$ promotes
formation of on-site bipolarons \cite{Hardikar07a}.  Importantly, the
co-operative interaction between the 2k$_F$ BOW and CDW in the BCDW
\cite{Mazumdar99a,Clay03a} implies that in the $V<V(U,S=0)$ region,
larger $g$ not only promotes the 4k$_F$ CO but also enhances the BCDW.
In our calculations in Fig.~\ref{fig-v2.25-a.27}(b) and (c) we we find
both these effects: a weak increase in $\chi_\rho(4k_F)$ at
intermediate T, and an even stronger increase in the $T\rightarrow 0$
values of $\chi_\rho(2k_F)$ and $\chi_B(2k_F)$. Actually, this result
is in qualitative agreement with our observation for the ground state
within the adiabatic limit of Eq.~(\ref{eqn-h}) that in the range
$0<V<V(U,S=0)$, $V$ enhances the BCDW \cite{Clay03a}.  The temperature
at which $\chi_\rho(2k_F)$ and $\chi_\rho(4k_F)$ cross does not change
significantly with larger $g$. Our results for $g=0.75$ for $V=2.75$
(not shown) are similar, except that the data are more noisy now
(probably because all parameters in Fig.~5 are much too large for the
larger $g$).  We conclude therefore that in the intermediate $V$
region, merely increasing $g$ does not change the BCDW nature of the
SP state. For $g$ to have a qualitatively different effect, $V$ should
be much closer to $V(U,S=0)$ or perhaps even larger.

\subsubsection{$V > V_c(U,S=0)$}

In principle, calculations of low temperature instabilities here
should be as straightforward as the weak intersite Coulomb interaction
$V < 2t$ regime. The 4k$_F$ CO - AFM1 state $\uparrow 0 \downarrow 0$
would occur naturally for the case of weak e-p coupling. Obtaining the
4k$_F$ CO-SP state $1=0=1\cdots0\cdots1$, with {\it realistic}
$V<\frac{1}{2}U$ is, however, difficult \cite{Clay03a}.  Previous work
\cite{Kuwabara03a}, for example, finds this state for
$V>\frac{1}{2}U$. Recent T-dependent mean-field calculations of Seo
{\it et al.} \cite{Seo07a} also fail to find this state for nonzero
$V$.  There are two reasons for this. First, the spin exchange here
involves charge-rich sites that are {\it second neighbors}, and are
hence necessarily small. Energy gained upon alternation of this weak
exchange interactions is even smaller, and thus the tendency to this
particular form of the SP transition is weak to begin with. Second,
this region of the parameter space involves either large $U$ (for
e.g., $U=10$, for which $V_c(U,S=0)\simeq 2$) or relatively large $V$
(for e.g. $V_c(U,S=0)\simeq 3$ for $U=8$). In either case such strong
Coulomb interactions make the applicability of mean-field theories
questionable.

Our QMC calculations do find the tendency to SP instability in this
parameter region.  In Fig.~\ref{fig-v3.5-a.24} we show QMC results for
$V>V_c(U=8,S=0)$. In contrast to Figs.~\ref{fig-v2.25-a.27} and
\ref{fig-v2.75-a.27}, $\chi_{\rho}(4k_F)$ now dominates over
$\chi_{\rho}(2k_F)$ at all T. The weaker peak at $q=2k_F$ at low T is
due to small differences in site charge populations between the
charge-poor sites of the Wigner crystal that arises upon bond
distortion \cite{Clay03a}, and that adds a small period 4 component to
the charge modulation.  The bond susceptibility $\chi_B(q)$ has a
strong peak at 2k$_F$ and a weaker peak at 4k$_F$, exactly as expected
for the 4k$_F$ CO-SP state.  Previous work has shown that the
difference in charge densities between the charge-rich and charge-poor
sites in the 4k$_F$ CO-SP ground state is considerably larger than in
the BCDW \cite{Clay03a}.

\subsection{Spin excitations from the BCDW}

The above susceptibility calculations indicate that in the
intermediate $V$ region (Figs.~\ref{fig-v2.25-a.27} and
\ref{fig-v2.75-a.27}) the BCDW ground state with $\cdots$1100$\cdots$
CO can evolve into the $\cdots$1010$\cdots$ CO as T increases. Our
earlier work had shown that for weak $V$, the BCDW evolves into the
4k$_F$ BOW at high T \cite{Ung94a,Clay03a}. Within the standard theory
of the SP transition \cite{Cross79a,Sorensen98a,Yu00a}, applicable to
the $\frac{1}{2}$-filled band, the SP state evolves into the
undistorted state for T $>$ T$_{SP}$. Thus the 4k$_F$ distortions, CO
and BOW, take the role of the undistorted state in the
$\frac{1}{4}$-filled band, and it appears paradoxical that the same
ground state can evolve into two different high T states. We show here
that this is intimately related to the nature of the spin excitations
in the $\frac{1}{4}$-filled BCDW. Spin excitations from the
conventional SP state leave the site charge occupancies unchanged. We
show below that not only do spin excitations from the
$\frac{1}{4}$-filled BCDW ground state lead to changes in the site
occupancies, {\it two different kinds of site occupancies are possible
in the localized defect states that characterize spin excited states
here.} We will refer to these as type I and type II defects, and
depending on which kind of defect dominates at high T (which in turn
depends on the relative magnitudes of $V$ and the e-p couplings), the
4k$_F$ state is either the CO or the BOW.

We will demonstrate the occurrence of type I and II defects in spin
excitations numerically.  Below we present a physical intuitive
explanation of this highly unusual behavior, based on a configuration
space picture. Very similar configuration space arguments can be found
elsewhere in our discussion of charge excitations from the BCDW in the
interacting $\frac{1}{4}$-filled band \cite{Clay01a}.

\begin{figure}[tb]
\centerline{\resizebox{3.0in}{!}{\includegraphics{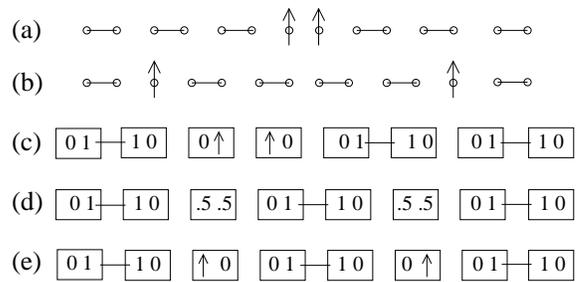}}}
\caption{(a) and (b) S = 1 VB diagrams in the Heisenberg SP chain. The
SP bond order in between the S = $\frac{1}{2}$ solitons in (b) is
opposite to that elsewhere.  (c) $\frac{1}{4}$-filled band equivalent
of (a); the singlet bonds in the background BCDW are between units
containing a pair of sites but a single electron (see text).  (d) and
(e) $\frac{1}{4}$-filled band equivalents of (b). Defect units with
two different charge distributions are possible.  The charge
distribution will depend on $V$.}
\label{fig-cartoon}
\end{figure}

We begin our discussion with the standard $\frac{1}{2}$-filled band,
for which the mechanism of the SP transition is well understood
\cite{Cross79a,Sorensen98a,Yu00a}.  Fig.~\ref{fig-cartoon}(a) shows in
valence bond (VB) representation the generation of a spin triplet from
the standard $\frac{1}{2}$-filled band.  Since the two phases of bond
alternation are isoenergetic, the two free spins can separate, and the
true wavefunction is dominated by VB diagrams as in
Fig.~\ref{fig-cartoon}(b), where the phase of the bond alternation in
between the two unpaired spins (spin solitons) is opposite to that in
the ground state.  With increasing temperature and increasing number
of spin excitations there occur many such regions with reversed bond
alternations, and overlaps between regions with different phases of
bond alternations leads ultimately to the uniform state.

The above picture needs to be modified for the dimerized dimer BCDW
state in the $\frac{1}{4}$-filled band, in which the single site of
the $\frac{1}{2}$-filled band system is replaced by a dimer unit with
site populations 1 and 0 (or 0 and 1), and the stronger interdimer
1--1 bond (weaker interdimer 0$\cdots$0 bond) corresponds to the
strong (weak) bond in the standard SP case.
Fig.~\ref{fig-cartoon}(c), showing triplet generation from the BCDW,
is the $\frac{1}{4}$-filled analog of Fig.~\ref{fig-cartoon}(a): a
singlet bond between the dimer units has been broken to generate a
localized triplet.
The effective repulsion between the free spins of
Fig.~\ref{fig-cartoon}(a) is due to the absence of binding between
them, and the same is expected in Fig.~\ref{fig-cartoon}(c). Because
the site occupancies {\it within} the dimer units are nonuniform now,
the repulsion between the spins is reduced from changes in {\it
intradimer} as well as interdimer site occupancies: the site
populations within the neighboring units can become uniform (0.5
each), or the site occupancies can revert to 1001 from 0110. There is
no equivalent of this step in the standard SP case.  The next steps in
the separation of the resultant defects are identical to that in
Fig.~\ref{fig-cartoon}(b), and we have shown these possible final
states in Fig.~\ref{fig-cartoon}(d) and Fig.~\ref{fig-cartoon}(e), for
the two different intraunit spin defect populations.  For
$V<V_c(U,S)$, defect units with site occupancies 0.5 occupancies (type
I defects) are expected to dominate; while for $V>V_c(U,S)$ site
populations of 10 and 01 (type II defects) dominate. From the
qualitative discussions it is difficult to predict whether the defects
are free, in which case they are solitons, or if they are bound, in
which case two of them constitute a triplet. What is more relevant is
that type I defects generate bond dimerization locally (recall that
the 4k$_F$ BOW has uniform site charge densities) while type II
defects generate local site populations 1010, which will have the
tendency to generate the 4k$_F$ CO.

The process by which the BCDW is reached from the 4k$_F$ BOW or the CO
as T is decreased from T$_{2k_F}$ is exactly opposite to the
discussion of spin excitations from the BCDW in
Fig.~\ref{fig-cartoon}. Now $\cdots$1100$\cdots$ domain walls appear
in the background bond-dimerized or charge-dimerized state.  In the
case of the $\cdots$1010$\cdots$ 4k$_F$ state, the driving force for
the creation of such a domain wall is the energy gained upon singlet
formation (for V $<$ V$_c$(U,S)).

Fig.~\ref{fig-cartoon}(e) suggests the appearance of localized
$\cdots$1010$\cdots$ regions in excitations from the
$\cdots$1100$\cdots$ BCDW SP state for $2|t| \leq V \leq V_c(U,S=0)$.
We have verified this with exact spin-dependent calculations for N =
16 and 20 periodic rings with adiabatic e-p couplings,
\begin{subequations}
\begin{eqnarray}
H^A_{SSH} &=& t\sum_i[1-\alpha(u_i-u_{i+1})](c_{i,\sigma}^{\dagger}c_{i+1,\sigma} + h.c.) \nonumber \\
&+& \frac{1}{2}K_S\sum_i(u_i-u_{i+1})^2 \label{eqn-adia-ssh}\\
H^A_{Hol} &=& g \sum_i v_in_i + \frac{1}{2}K_H\sum_i v_i^2 \label{eqn-adia-hol}
\end{eqnarray}
\end{subequations}
and $H_{ee}$ as in Eq.~(\ref{eqn-uv}). In the above $u_i$ is the
displacement from equilibrium of a molecular unit and $v_i$ is a
molecular mode.  Note that due to the small sizes of the systems we
are able to solve exactly, the e-p couplings in Eq.~(\ref{eqn-adia-ssh})
and Eq.~(\ref{eqn-adia-hol}) cannot be directly compared to those in
Eq.~(\ref{eqn-h}). 
In the present case, we take the smallest e-p couplings necessary to
generate the BCDW ground state self-consistently within the adiabatic
Hamiltonian \cite{Clay03a,Riera00a,Riera01a}. We then determine the
bond distortions, site charge occupancies and spin-spin correlations
self-consistently with the same parameters for excited states with
$S_z>0$.  
\begin{figure}[tb]
\centerline{\resizebox{3.0in}{!}{\includegraphics{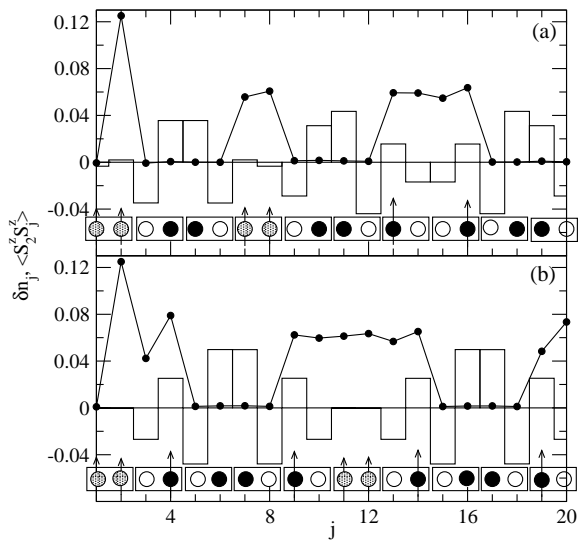}}}
\caption{(a) Self-consistent site charges (vertical bars) and spin-spin
correlations (points, lines are to guide the eye) in the N=20 periodic
ring with $U=8$, $V=2.75$, $\alpha^2/K_S=1.5$, $g^2/K_H=0.32$ for $S_z=2$.
(b) Same parameters, except $S_z=3$. Black and white circles at the bottoms
of the panels denote charge-rich and charge-poor sites, respectively, while
gray circles denote sites with population nearly exactly 0.5. The arrows
indicate locations of the defect units with nonzero spin densities.}
\label{fig-exact}
\end{figure}

In Fig.~\ref{fig-exact}(a) we have plotted the deviations of the site
charge populations from average density of 0.5 for the lowest $S_z$=2
state for N = 20, for $U=8$, $V=2.75$ and e-p couplings as given in
the figure caption.  Deviations of the charge occupancies from the
perfect $\cdots$0110$\cdots$ sequence identify the defect centers, as
seen from Fig.~\ref{fig-cartoon}(c) - (e).  In the following we refer
to dimer units composed of sites i and j as [i,j].  Based on the
charge occupancies in the Fig, we identify units [1,2], [7,8], [13,14]
and [14,15] as the defect centers.  Furthermore, based on site
populations of nearly exactly 0.5, we identify defects on units [1,2]
and [7,8] as type I; the populations on units [13,14] and [14,15]
identify them type II. Type I defects appear to be free and soliton
like, while type II defects appear to be bound into a triplet state,
but both could be finite size effects.

Fig.~\ref{fig-cartoon}(d) and (e) suggest that the spin-spin
z-component correlations, $\langle S_i^zS_j^z \rangle$, are large and
positive {\it only between pairs of sites which belong to the defect
centers}, as all other spins are singlet-coupled. For our
characterization of units [1,2], [7,8], [13,14] and [14,15] to be
correct therefore, $\langle S_i^zS_j^z \rangle$ must be large and
positive if sites $i$ and $j$ both belong to this set of sites, and
small (close to zero) when either $i$ or $j$ does not belong to this
set (as all sites that do not belong to the set are singlet-bonded).
We have superimposed the calculated z-component spin-spin correlations
between site 2 and all sites j = 1 -- 20.  The spin-spin correlations
are in complete agreement with our characterization of units [1,2],
[7,8], [13,14] and [14,15] as defect centers with free spins.  The
singlet 1--1 bonds between nearest neighbor charge-rich sites in
Fig.~\ref{fig-cartoon}(c) - (e) require that spin-spin couplings
between such pairs of sites are large and negative, while spin-spin
correlations between one member of the pair any other site is
small. We have verified such strong spin-singlet bonds between sites 4
and 5, 10 and 11, and 18 and 19, respectively (not shown). Thus
multiple considerations lead to the identification of the same sites
as defect centers, and to their characterization as types I and II.

We report similar calculations in Fig.~\ref{fig-exact}(b) for
S$_z=3$. Based on charge occupancies, four out of the six defect units
with unpaired spins in the S$_z$ = 3 state are type II; these occupy
units [3,4], [9,10], [13,14] and [19,20]. Type I defects occur on
units [1,2] and [11,12]. As indicated in the figure, spin-spin
correlations are again large between site 2 and all other sites that
belong to this set, while they are again close to zero when the second
site is not a defect site. As in the previous case, we have verified
singlet spin couplings between nearest neighbor pairs of charge-rich
sites. There occur then exact correspondences between charge densities
and spin-spin correlations, exactly as for $S_z=2$.

The susceptibility calculations in Fig.~\ref{fig-v2.25-a.27} and
Fig.~\ref{fig-v2.75-a.27} are consistent with the microscopic
calculations of spin defects presented above. As 4k$_F$ defects are
added to the 2k$_F$ background, the 2k$_F$ susceptibility peak is
expected to broaden and shift towards higher $q$. This is exactly what
is seen in the charge susceptibility, Fig.~\ref{fig-v2.25-a.27}(a) and
Fig.~\ref{fig-v2.75-a.27}(a) as T is increased. A similar broadening
and shift is seen in the bond order susceptibility as well.

\section{Discussions and Conclusions}

In summary, the SP state in the $\frac{1}{4}$-filled band CTS is
unique for a wide range of realistic Coulomb interactions. Even when
the 4k$_F$ state is the Wigner crystal, the SP state can be the
$\cdots$1100$\cdots$ BOW. For $U=8$, for example, the transition found
here will occur for $2<V<3$. This novel T-dependent transition from
the Wigner crystal to the BCDW is a consequence of the spin-dependence
of $V_c$. Only for $V>3$ here can the SP phase be the tetramerized
Wigner crystal $1=0=1\cdots0\cdots1$ (note, however, that $V \leq 4$
for $U=8$). We have ignored the intrinsic dimerization along the 1D
stacks in our reported results, but this increases $V_c(U,S=0)$ even
further, and makes the Wigner crystal that much more unlikely
\cite{Shibata01a}.  Although even larger $U$ ($U=10$, for example)
reduces $V_c(U,S=0)$, we believe that the Coulomb interactions in the
(TMTTF)$_2$X lie in the intermediate range.

A Wigner crystal to BCDW transition would explain most of the
experimental surprises discussed in Section II. The discovery of the
charge redistribution upon entering the SP phase
\cite{Fujiyama06a,Nakamura07a} in X = AsF$_6$ and PF$_6$ is probably
the most dramatic illustration of this. Had the SP state maintained
the same charge modulation pattern as the Wigner crystal that exist
above T$_{2k_F}$, the difference in charge densities between the
charge-rich and the charge-poor sites would have changed very slightly
\cite{Clay03a}. The dominant effect of the SP transition leading to
$1=0=1\cdots0\cdots1$ is only on the charge-poor sites, which are now
inequivalent (note that the charge-rich sites remain equivalent). The
difference in charge density of the charge-rich sites and the average
of the charge density of the charge-poor sites thus remains the same
\cite{Clay03a}.  In contrast, the difference in charge densities
between the charge-rich and the charge-poor sites in the BCDW is
considerably smaller than in the Wigner crystal \cite{Clay03a}, and we
believe that the experimentally observed smaller charge density
difference in the SP phase simply reflects its BCDW character.

The experiments of references \onlinecite{Fujiyama06a,Nakamura07a}
should not be taken in isolation: we ascribe the competition between
the CO and the SP states in X = PF$_6$ and AsF$_6$, as reflected in
the different pressure-dependences of these states
\cite{Zamborszky02a}, to their having different site charge
occupancies. The observation that the charge density difference in X =
SbF$_6$ decreases considerably upon entering the SP phase from the
AFM1 phase \cite{Yu04a} can also be understood if the AFM1 and the SP
phases are assigned to be Wigner crystal and the BCDW, respectively.
The correlation between larger T$_{CO}$ and smaller T$_{SP}$
\cite{Pouget06a} is expected. Larger T$_{CO}$ implies larger effective
$V$, which would lower T$_{SP}$. This is confirmed from comparing
Figs.~4 and 5: the temperature at which $\chi_{\rho(2k_F)}$ begins to
dominate over $\chi_{\rho(4k_F)}$ is considerably larger in Fig.~4
(smaller $V$) than in Fig.~5 (larger $V$). The isotope effect, strong
enhancement of the T$_{CO}$ (from 69 K to 90 K) with deuteration of
the methyl groups in X = PF$_6$, and concomitant decrease in T$_{SP}$
\cite{Nad05a,Furukawa05a} are explained along the same
line. Deuteration decreases $\omega_H$ in Eq.~(\ref{eqn-h}), which has
the same effect as increasing $V$. Thus from several different
considerations we come to the conclusion that the transition from the
Wigner crystal to the BCDW that we have found here theoretically for
intermediate $V/|t|$ does actually occur in (TMTTF)$_2$X that undergo
SP transition. This should not be surprising.  Given that the 1:2
anionic CTS lie in the ``weak'' $V/|t|$ regime, TMTTF with only
slightly smaller $|t|$ (but presumably very similar $V$, since
intrastack intermolecular distances are comparable in the two
families) lies in the ``intermediate'' as opposed to ``strong''
$V/|t|$ regime.

Within our theory, the two different antiferromagnetic regions that
straddle the SP phase in Fig.~1, AFM1 and AFM2, have different charge
occupancies. The Wigner crystal character of the AFM1 region is
revealed from the similar behavior of T$_N$ and T$_{CO}$ in
(TMTTF)$_2$SbF$_6$ under pressure \cite{Yu04a}, indicating the absence
of the competition of the type that exists between CO and SP, in
agreement with our assignment. The occurrence of a Wigner crystal AFM1
instead of SP does not necessarily imply a larger $V/|t|$ in the
SbF$_6$. A more likely reason is that the interaction with the
counterions is strong here, and this interaction together with $V$
pins the electrons on alternate sites. (TMTSF)$_2$X, and possibly
(TMTTF)$_2$Br, belong to the AFM2 region. The observation that the CDW
and the spin-density wave in the TMTSF have the same periodicities
\cite{Pouget96a} had led to the conclusion that the charge occupancy
here is $\cdots$1100$\cdots$ \cite{Mazumdar99a}. This conclusion
remains unchanged. Finally, we comment that the observation of Wigner
crystal CO \cite{Kanoda} in (DI-DCNQI)$_2$Ag is not against our
theory, as the low T phase here is antiferromagnetic and not SP. We
predict the SP system (DMe-DCNQI)$_2$Ag to have the
$\cdots$1100$\cdots$ charge ordering.

In summary, it appears that the key concept of spin-dependent $V_c$
within Eq.~(\ref{eqn-h}) can resolve most of the mysteries associated
with the temperature dependence of the broken symmetries in the
$\frac{1}{4}$-filled band CTS. One interesting feature of our work
involves demonstration of spin excitations from the BCDW state that
necessarily lead to local changes in site charges.  Even for weak
Coulomb interactions, the $\frac{1}{4}$-filled band has the BCDW
character \cite{Ung94a}. The effects of magnetic field on
$\frac{1}{4}$-filled band CDWs is of strong recent interest
\cite{Graf04a,McDonald04a}. We are currently investigating the
consequences of the Zeeman interaction on the mixed spin-charge
excitations of the BCDW.

\section{Acknowledgments}

We acknowledge illuminating discussions with S.E. Brown.
This work was supported by the Department of Energy grant DE-FG02-06ER46315
and the American Chemical Society Petroleum Research Fund.


\begin{thebibliography}{75}
\bibitem[{foo()}]{footnote}
\bibinfo{note}{We assume equal intermolecular distances along the organic
  stack. We ignore the intrinsic dimerization that can exist in TMTTF stacks
  and that can lead to charge localization above \protect{T$_{CO}$}}.
\end{thebibliography}
\end{document}